\documentclass{article}
\usepackage{amsmath,graphicx}
\usepackage{cite}
\usepackage{amssymb,amsfonts,mathrsfs}
\usepackage{textcomp}
\usepackage{xcolor}
\usepackage{xspace}
\usepackage{hyperref, booktabs}
\usepackage[ruled,vlined,noend]{algorithm2e}
\usepackage{caption}
\usepackage{subcaption}
\usepackage{lipsum}
\usepackage{multicol, multirow}
\usepackage{makecell}

\DeclareMathOperator*{\argmax}{arg\,max}

\makeatletter
\DeclareRobustCommand\onedot{\futurelet\@let@token\@onedot}
\def\@onedot{\ifx\@let@token.\else.\null\fi\xspace}
\def\eg{\emph{e.g}\onedot} 
\def\ie{\emph{i.e}\onedot}

\makeatother

\title{Test-Time Detection of Backdoor Triggers \\
for Poisoned Deep Neural Networks}

\author{Xi Li, Zhen Xiang, David J. Miller, George Kesidis\\
School of EECS, Pennsylvania State University\\
\{xzl45,djm25,gik2\}@psu.edu}
\begin{document}
\maketitle

\begin{abstract}
Backdoor (Trojan) attacks are emerging threats against deep neural networks (DNN). A DNN being attacked will predict to an attacker-desired {\it target class} whenever a test sample from any {\it source class} is embedded with a {\it backdoor pattern}; while correctly classifying clean (attack-free) test samples.
Existing backdoor defenses have shown success in detecting whether a DNN is attacked and in reverse-engineering the backdoor pattern in a ``post-training'' regime: the defender has access to the DNN to be inspected and a small, clean dataset collected independently, but has no access to the (possibly poisoned) training set of the DNN. 
However, these defenses neither catch culprits in the act of triggering the backdoor mapping, nor mitigate the backdoor attack at test-time. In this paper, we propose an ``in-flight" defense against backdoor attacks on image classification that 1) detects use of a backdoor trigger at {\it test-time}; and 2) infers the class of origin (source class) for a detected trigger example. The effectiveness of our defense is demonstrated experimentally against different strong backdoor attacks.
\end{abstract}

\section{Introduction}
\label{sec:intro}
Deep neural networks (DNN) have shown impressive performance in many applications, but are vulnerable to adversarial attacks \cite{Proceedings}. Recently, backdoor attacks have been successfully launched against DNNs for image classification\cite{BadNets,Targeted-Backdoor,Trojan,DBLP:journals/corr/abs-1909-02742,HiddenTrigger}, speech recognition \cite{Trojan}, text classification \cite{8836465}, 3D point cloud classification\cite{ZhenICCV}, and deep regression \cite{LKML21}. 
Typically, a backdoor attack is launched by embedding a specific backdoor pattern into a small number of training samples from one or more source classes and (mis)labeling them to an attacker-desired target class. 
The DNN being attacked will classify to the target class whenever a test sample originally from the source classes is embedded with the same backdoor pattern; while still correctly classifying clean test samples, {\it i.e.} without the backdoor pattern \cite{BadNets, YimingLi}. 

Early backdoor defenses aim to inspect the training set and detect samples embedded with the backdoor pattern \cite{SS, AC}. But they are not applicable to scenarios where the training set of the DNN is not available  (\eg, legacy or proprietary systems); thus they are not discussed here. A more practical {\it post-training} scenario assumes that the defender, \eg, a downstream app user, has {\it no access} to the DNN's training set. Defenses for this scenario detect whether a trained DNN is backdoor attacked, infer the target class if an attack is detected, and usually reverse-engineer the backdoor pattern used by the attacker \cite{TNNLS,NC,B3D,TABOR,DBLP:conf/icassp/XiangMK20}. A post-training defender does possess a small, clean dataset collected independently -- this dataset is not sufficient for training a clean (backdoor-free) DNN \textit{from scratch} if an attack is detected.

However, existing post-training defenses cannot catch entities in the act of exploiting the backdoor mapping at test-time. In particular, while these defenses successfully detect backdoor attacks and correctly infer the attacker's target class, they neither decide whether a test sample classified to the target class is backdoor-free nor mitigate the attack by correctly classifying test samples embedded with the backdoor pattern to their true source classes (as a clean DNN will do).

In this paper, we make the following contributions:
(1) We propose a method that, at {\it test-time}, detects image backdoor triggers and infers the source class for detected backdoor trigger images, given the reverse-engineered backdoor pattern and the target class obtained from a post-training backdoor detector. While we focus on image classification here, our method can be extended to other domains.
(2) Our detector requires neither access to the DNN's training set nor any DNN training/fine-tuning. It is efficient in data used for detection and computational resources.
(3) We show the effectiveness of our detector experimentally for a wide variety of DNN architectures, datasets, and backdoor attack configurations.

\section{Related Work}
\label{sec:RW}
Neural Cleanse (NC) \cite{NC} detects test images embedded with backdoor triggers by their activations on neurons that are most relevant to the (estimated) backdoor. If the input image has activations higher than a given threshold on those neurons, it is deemed a backdoor trigger image. Its performance highly depends on the choice of abnormal neurons and detection threshold.
\cite{B3D} proposes a black-box backdoor detection (B3D), which captures backdoor triggers based on the difference in model outputs for a test image with and without the reverse-engineered backdoor pattern embedded. The two model outputs are hypothesized to be very different for a clean input but very similar for a backdoored input. 
However, for both backdoor-trigger images classified to the target class and clean target class images, embedding a backdoor trigger is expected to have little impact on the model outputs, thus they cannot be effectively distinguished by B3D.
A STRong Intentional Perturbation (STRIP) method proposed in \cite{STRIP} linearly blends test images with a few clean images and detects backdoor triggers based on the average entropy of model posteriors of these blended images. A small entropy indicates a backdoor-trigger input.  However, it is sensitive to model complexity, as will be shown in Sec.\ref{sec:exp}.

For inferring the class of origin for a detected backdoor-trigger image, NC \cite{NC} proposes to patch the poisoned DNN by fine-tuning the DNN on 10\% of the original (backdoor-free) training set, 20\% of which are embedded with the reverse engineered backdoor pattern and correctly labeled. However, it is unreasonable to assume the defender has access to the clean training set, which is inconsistent with the post-training scenario.


\section{Threat Model and Defense Assumptions}
\label{sec:threat_model}

{\bf Classification domain:} Like most existing works, we focus on image classification for simplicity; our method is easily extended to domains other than images.\\
{\bf Attacker's goals:} An attacker aims to have the DNN learn to classify to a target class $t\in\mathcal{C}$, whenever a test image from any source class $c\in\mathcal{S}_A\subseteq \mathcal{C}\setminus\{t\}$ is embedded with a specific backdoor pattern; while not degrading the DNN's accuracy on backdoor-free test images. Here, $\mathcal{C}$ is the category space of the classification domain.\\
{\bf Attacker's knowledge:} The attacker has full knowledge of $\mathcal{C}$ and the ability to collect valid images from source classes $\mathcal{S}_A$. But the attacker has no specific knowledge about any defenses that may be deployed to detect the attack.\\
{\bf Attack strategy:} We consider classical backdoor attacks launched by poisoning the DNN's training set \cite{BadNets,Targeted-Backdoor}. The attacker embeds the same backdoor pattern that will be used at test-time into a small set of images collected from $\mathcal{S}_A$, (mis)labels them to the target class $t$, and injects them into the DNN's training set.
The backdoor pattern can be an imperceptible additive perturbation bounded by its $L_p$ norm\cite{TNNLS}, or a perceptible (but hopefully scene-plausible) small patch embedded in (or blended with) an image\cite{BadNets,Targeted-Backdoor,Trojan}.\\
{\bf Defender's goals:} Given any \textit{single} test image predicted to the target class $t$ inferred by a post-training defense, our in-flight detector \textit{aims} to: 1) detect whether the image contains the backdoor pattern (\ie, a backdoor trigger), and, if so, 2) infer its true class of origin (source class). \\
{\bf Defender's knowledge:} The defender has access to: 1) the DNN detected as attacked, together with the target class $t$\footnote{We only consider the cases where the post-training defender successfully detects the backdoor attack with correct inference of the target class.} inferred by the post-training defense \cite{NC,TNNLS}; and 2) a (small) set of clean images from all classes of $\mathcal{C}$. The defender does require the backdoor pattern estimated by the same post-training defense, but does not make any assumptions about the type, shape, or location of the attacker's backdoor pattern. Thus, our defense can be coupled with any post-training defense and addresses a variety of backdoor patterns. Finally, our defender has no knowledge of the source classes involved in the attack, nor enough legitimate data or computational resources to train a clean DNN \textit{from scratch}.

\section{Methodology}\label{sec:method}

\subsection{Notation}

We denote the DNN (for which a post-training defense has detected an attack) by $f: \mathcal{X}\rightarrow \mathcal{C}$, where $\mathcal{X}$ is the input (image) space and $\mathcal{C}$ is the label space. We denote $f_L$ as the output of any internal layer $L$. Let $t$ be the detected target class. Then, the actual source classes $\mathcal{S}_A$ involved in the attack, unknown to our defender, is a subset of $\mathcal{S}=\mathcal{C}\setminus\{t\}$, i.e., all classes except the target class. We define $\mathcal{D_\text{Defense}}=\bigcup_{c\in\mathcal{C}}\mathcal{D}_c$ as the clean dataset possessed by the defender, where $\mathcal{D}_c$ contains images labeled to class $c$. Backdoor patterns of different types may be crafted and embedded in images in very differently ways. For simplicity, we use a universal notation $\Delta$ to denote a backdoor pattern irrespective of its type. Moreover, we define the embedding function associated with $\Delta$ as $g(\cdot, \Delta):\mathcal{X}\rightarrow \mathcal{X}$, such that a clean image $\boldsymbol{x}\in{\mathcal X}$ embedded with $\Delta$ can be written as $\tilde{\boldsymbol{x}} = g(\boldsymbol{x}, \Delta)$. Finally, we denote the backdoor pattern estimated by the post-training defense as $\widehat{\Delta}$.

\subsection{In-flight Backdoor Defense}\label{sec:dect}

The estimated backdoor pattern $\widehat{\Delta}$ obtained by an {\it effective} post-training defense will likely elicit the same targeted misclassification to the backdoor target class $t$ as the true backdoor pattern $\Delta$ (used by the attacker and unknown to the defender) when they are embedded in source class images (as will be shown by our experiments). But these two patterns may have very different intensity values in the image space, as visualized by \cite{NC}. Therefore, it is unreliable to determine whether a test image classified to class $t$ is embedded with $\Delta$ by directly looking for the estimated $\widehat{\Delta}$ in the image. Even if a test image embedded with $\Delta$ is successfully detected, directly removing an estimated $\widehat{\Delta}$ (\eg, subtracting the estimated additive perturbation pattern \cite{TNNLS}) from the image will not likely remove the true backdoor pattern $\Delta$ completely -- the image may still be classified to the target class $t$.

However, in deep layers close to the DNN output (\eg, the penultimate layer), the true backdoor pattern $\Delta$ and its empirical estimation $\widehat{\Delta}$ will likely activate the {\it same} set of neurons when embedded in images from the same source class. These neurons are trained (on the poisoned training set) to activate for the backdoor mapping, and are likely {\it different} from the neurons activating for {\it typical} target class images. Thus, if a test image classified to the target class $t$ is actually an image from some source class embedded with the backdoor pattern $\Delta$, its deep layer activations are expected to be: a) similar to the activations for most images from the same source class embedded with the estimated pattern $\widehat{\Delta}$, and b) different from the activations for typical images from class $t$.

Based on the above intuition, our in-flight backdoor detection steps are as follows: First, for each non-target class $\forall c\in\mathcal{S}$, we embed the estimated backdoor pattern $\widehat{\Delta}$ in the clean images used for detection and get $\tilde{\mathcal{D}}_c=\{g(\boldsymbol{x}, \widehat{\Delta})| \boldsymbol{x}\in \mathcal{D}_c\}$. Then we feed all images in $\bigcup_{c\in\mathcal{S}}\tilde{\mathcal{D}}_c$ and the clean target class images $\mathcal{D}_t$ to the DNN being attacked to get their deep layer features $\mathcal{Z}_c = \{f_L(\tilde{\boldsymbol{x}})| \tilde{\boldsymbol{x}}\in \tilde{\mathcal{D}}_c\}$, $\forall c\in\mathcal{S}$ and $\mathcal{Z}_t= \{f_L(\boldsymbol{x})| \boldsymbol{x}\in \mathcal{D}_t\}$, respectively. All the features in $\mathcal{Z}=\bigcup_{c\in\mathcal{C}}\mathcal{Z}_c$ are then standardized to have mean 0 and standard deviation 1. For each class $c\in\mathcal{C}$, we learn a density model (\eg, Gaussian mixture model) with parameters $\boldsymbol{\theta}_c = \argmax_{\boldsymbol{\theta}}\Pi_{\boldsymbol{z}\in\mathcal{Z}_c}P[\boldsymbol{z}|\boldsymbol{\theta}]$. {\it At test-time}, for any test image $\boldsymbol{w}$ with $f(\boldsymbol{w})=t$, we measure its likelihood $\mathcal{L}_c = P[f_L(\boldsymbol{w})|\boldsymbol{\theta}_c]$ under the density model for each $c\in\mathcal{C}$. If $\argmax_{c\in\mathcal{C}}\mathcal{L}_c\neq t$, $\boldsymbol{w}$ is deemed to contain the backdoor pattern. We then reject the prediction of the DNN and infer that it is originally from class $s = \argmax_{c\in\mathcal{C}}\mathcal{L}_c$. Otherwise, $\boldsymbol{w}$ is deemed clean and its class prediction $t$ is accepted.

Our method has the following advantages over existing ones. Unlike \cite{NC,B3D,STRIP} which are sensitive to the choice of hyper-parameters, \eg, a detection threshold, we do not require careful choices of hyper-parameters. Besides, our method only needs relatively few clean images (100 images per class in our experiments) for detection, and is computationally cheap, as it does not involve tuning a complicated DNN for example. 
Note that the most time-consuming part of our defense -- learning class-conditional density models -- is done offline.

\section{Experiments}
\label{sec:exp}
\subsection{Experiment Setup} \label{sec:exp_setup}
\textbf{Dataset}: We mainly use the benchmark CIFAR-10 dataset, which contains 60k $32\times32$ color images from 10 classes, with 5k images per class for training and 1k images per class for testing \cite{cifar10}. Experiments on other datasets including MNIST\cite{MNIST}, PubFig\cite{PubFig}, and F-MNIST\cite{FMNIST} are in Sec. \ref{sec:other_datasets}. \\
\textbf{Data allocation}: Following the assumptions in Sec. \ref{sec:threat_model}, we randomly split the test set of CIFAR-10 into $\mathcal{D}_{\text{Defense}}$ and $\mathcal{D}_{\text{Test}}$, where $\mathcal{D}_{\text{Defense}}$ consists of 100 images per class, and $\mathcal{D}_{\text{Test}}$ is used for performance evaluation. \\
\textbf{Attack settings}: We consider typical backdoor attacks described in Sec. \ref{sec:threat_model}. We arbitrarily choose class 9 as the target class for all attacks. We consider the following three backdoor patterns: 1) an additive perturbation that looks like a ``chess board'' (\textbf{CB}) used in \cite{TNNLS}; 2) a single pixel set to 255 (\textbf{SP}) \cite{MLSP}; and 3) a $3\times3$ white box (\textbf{WB}) embedded in the bottom right of an image\cite{BadNets}. For each type of backdoor pattern, we create two attacks by: 1) embedding the backdoor pattern in 1000 training images randomly selected from class 0 (single class attack); 2) embedding the backdoor pattern in 100 training images randomly selected from each class, except for the target class (multi-class attack). \\
\textbf{Training settings}: For each attack, we train a DNN with the widely used ResNet-18\cite{ResNet} architecture on the backdoor poisoned CIFAR-10 training set. Training is performed for 150 epochs with learning rate 0.1 (reduced by 0.5 per 50 epochs) and batch size 32.\\
\textbf{Defense settings}: 
For each attack, we first apply a post-training detector to the DNN being attacked to infer the target class $t$ and reverse-engineer its corresponding backdoor pattern $\widehat{\Delta}$, using $\mathcal{D}_{\text{Defense}}$ reserved for detection.
In particular, we apply the detector in \cite{TNNLS} to attacks with backdoor pattern CB since it is an additive perturbation; and the detector in \cite{NC} to attacks with backdoor patterns SP or WB as they are small patches (not additive perturbations) embedded in an image.
For our in-flight detector, we consider Gaussian mixture models and choose the penultimate layer features for (maximum likelihood) density model estimation.

\subsection{Main Experimental Results on CIFAR-10} 

First, we show that the attacks we created are sufficiently effective for thoroughly evaluating detection performance. Such effectiveness is evaluated by: 1) the attack success rate (\textbf{ASR}) defined as the fraction of clean images from the source classes in $\mathcal{D}_{\text{Test}}$ that are misclassified to the target class when the ground truth (\textbf{GT}) backdoor pattern is embedded; and 2) the clean test accuracy (\textbf{ACC}) defined as the DNN's accuracy on $\mathcal{D}_{\text{Test}}$. As shown in Table~\ref{tab:1}, all attacks have high ASR and almost no degradation in ACC compared with the baseline ACC of a DNN trained with no backdoor. 
Second, we show the ASR for each attack with the backdoor pattern reverse-engineered (\textbf{RE}) by the post-training defense instead of the GT pattern in Table~\ref{tab:1}. The RE patterns induce similarly high misclassification rate to the target class, when embedded in clean source class images from $\mathcal{D}_{\text{Test}}$, as GT patterns.

\begin{table}[t!] 
\centering
\small
\begin{tabular}{lcccc}
\toprule
\multirow{2}{*}{Attack pattern} & \multicolumn{2}{c}{Single class attack} & \multicolumn{2}{c}{Multi-class attack} \\
\cline{2-5}
& ACC & ASR & ACC & ASR \\
\hline
No attack & 0.9387 & NA     & 0.9387 & NA \\
CB-GT     & 0.9360 & 0.9955 & 0.9381 & 0.9954 \\
CB-RE     & NA     & 1.0000 & NA     & 0.9876 \\
SP-GT     & 0.9354 & 0.9488 & 0.9337 & 0.9565 \\
SP-RE     & NA     & 0.9900 & NA     & 0.9953 \\
WB-GT     & 0.9324 & 0.9411 & 0.9340 & 0.9497 \\
WB-RE     & NA     & 0.9970 & NA     & 0.9354 \\
\bottomrule
\end{tabular}
\caption{ASR and ACC for attacks using GT patterns; and ASR for the RE patterns obtained by post-training defenses applied to these attacks. ``NA'' represents ``not applicable''.}
\label{tab:1}
\end{table}

We evaluate the effectiveness of our detector in comparison with three other in-flight detectors NC\cite{NC}, B3D\cite{B3D}, and STRIP\cite{STRIP}.
The metrics for performance evaluation include true positive rates (\textbf{TPR}, \ie, the fraction of backdoor-trigger images correctly detected), false positive rates (\textbf{FPR}, \ie, the fraction of clean images falsely detected) and source class inference accuracies (\textbf{SIA}, \ie, the fraction of backdoor-trigger images with correct inference of the source class).
As shown in Table~\ref{tab:2}, for all attacks, our detector performs perfectly -- almost all the backdoor-trigger images are correctly identified, and no clean target class images are falsely reported. 
Following \cite{NC, B3D, STRIP}, we build in-flight detectors for NC, B3D, and STRIP. As they do not announce the thresholds for detection, we test them with various thresholds and exhibit the best TPR at an FPR of around 5\%\footnote{A 5\% FPR is not achievable by B3D for some attacks (even as the detection threshold is varied over a wide range.}. For STRIP, we set the weight of the incoming input as 0.5 in image blending. 
NC does not correctly detect any backdoor-trigger images in a single class attack using the pattern WB.
B3D does not perform well under all the attacks, as embedding the estimated backdoor pattern has little impact on the model outputs for both backdoor-trigger images and clean target class images. STRIP does not perform well either (for ResNet18 and CIFAR-10) -- STRIP is more effective for the DNN architectures, datasets, and attack configurations used in \cite{STRIP}.

\begin{table}[t!] 
\centering
\small
\begin{tabular}{cp{0.78cm}p{0.78cm}p{0.78cm}p{0.78cm}p{0.78cm}p{0.78cm}}
\toprule
\multirow{2}{*}{\makecell[c]{Attack \\ pattern}} & \multicolumn{3}{c}{Single class attack} & \multicolumn{3}{c}{Multi-class attack} \\
\cline{2-7}
& TPR & FPR & SIA & TPR & FPR & SIA \\
\hline
\multicolumn{7}{c}{Likelihood-based in-flight backdoor defender} \\
\hline
CB & 0.9922 & 0.0 & 0.8392 & 0.9997 & 0.0 & 0.6946 \\
SP & 0.9813 & 0.0 & 0.7728 & 0.9454 & 0.0 & 0.632 \\
WB & 0.9847 & 0.0 & 0.8607 & 0.9992 & 0.0 & 0.8945 \\
\hline
\multicolumn{7}{c}{NC} \\
\hline
CB & 0.9855 & 0.0488 & 0.9444 & 0.9962 & 0.0533 & 0.8765 \\
SP & 0.8088 & 0.0544 & 0.8833 & 0.9043 & 0.0511 & 0.8963 \\
WB & 0.0    & 0.0522 & 0.8667 & 0.8644 & 0.0522 & 0.8086 \\
\hline
\multicolumn{7}{c}{B3D} \\
\hline
CB & 0.0788 & 0.0511 & NA & 0.9872 & 0.9955 & NA \\
SP & 0.5333 & 0.1066 & NA & 0.1814 & 0.0522 & NA \\
WB & 0.0011 & 0.0500 & NA & 0.0535 & 0.0511 & NA \\
\hline
\multicolumn{7}{c}{STRIP} \\
\hline
CB & 0.0822 & 0.0533 & NA & 0.0218 & 0.0555 & NA \\
SP & 0.1333 & 0.0588 & NA & 0.1555 & 0.0588 & NA \\
WB & 0.0088 & 0.0588 & NA & 0.0011 & 0.0633 & NA \\
\bottomrule
\end{tabular}
\caption{TPR, FPR and SIA for our defense, compared with three other in-flight defenses, NC, B3D, and STRIP, against all the created attacks. ``NA'' signifies ``not applicable''.}
\label{tab:2}
\end{table}

Since the NC paper does not mention the learning rate used for fine-tuning the DNN, in our experiments, we fine-tune the poisoned DNN using various learning rates. From Table~\ref{tab:3}, the severe fluctuations in SIAs of models poisoned by the multi-class attack using pattern WB shows the sensitivity of NC to the learning rate. We thus apply NC with all the learning rates, reporting the {\it best} SIA in Table~\ref{tab:2}.
Our defender performs relatively well in inferring source classes for the detected backdoor trigger images, though it is not as good as the \textit{best results} of NC. However 
note that our method requires neither a clean set as large as the one used by NC for fine-tuning, nor careful choices of hyper-parameters.

\begin{table}[t!] 
\centering
\small
\begin{tabular}{lllllll}
\toprule
LR & 10 & 1 & $10^{-1}$ & $10^{-2}$ & $10^{-3}$ & $10^{-4}$ \\
\hline
SIA & 0.1114 & 0.4291 & 0.6033 & 0.8086 & 0.7103 & 0.6348 \\
\bottomrule
\end{tabular}
\caption{SIA of NC fluctuates with the learning rate (LR) used for DNN fine-tuning.}
\label{tab:3}
\end{table}

We also implement our detector on the last 4 convolutional layers of ResNet-18 -- the internal layers just before the penultimate layer. We reduce the dimensionality of the internal layer activations to that of the penultimate layer activations via, \eg, average pooling. As shown in Table~\ref{tab:4}, the nearly perfect TPRs and FPRs demonstrate that the choice of the deep layer has little impact on our in-flight backdoor trigger detection.
However, the choice of the deep layer affects source class inference for backdoor-trigger images. Our method achieves the best SIA on the penultimate layer. 

\begin{table}[t!] 
\centering
\small
\begin{tabular}{cp{0.78cm}p{0.78cm}p{0.78cm}p{0.78cm}p{0.78cm}p{0.78cm}}
\toprule
\multirow{2}{*}{\makecell[c]{Attack \\ pattern}} & \multicolumn{3}{c}{Single class attack} & \multicolumn{3}{c}{Multi-class attack} \\
\cline{2-7}
& TPR & FPR & SIA & TPR & FPR & SIA \\
\hline
\multicolumn{7}{c}{Penultimate layer} \\
\hline
CB & 0.9922 & 0.0 & 0.8392 & 0.9997 & 0.0 & 0.6946 \\
SP & 0.9813 & 0.0 & 0.7728 & 0.9454 & 0.0 & 0.632 \\
WB & 0.9847 & 0.0 & 0.8607 & 0.9992 & 0.0 & 0.8945 \\
\hline
\multicolumn{7}{c}{1st convolutional layer from the last} \\
\hline
CB & 0.9977 & 0.0 & 0.7897 & 0.9996 & 0.0 & 0.6488 \\
SP & 0.9894 & 0.0 & 0.7562 & 0.9807 & 0.0 & 0.5271 \\
WB & 0.9941 & 0.0057 & 0.8598 & 0.9997 & 0.0034 & 0.8638 \\
\hline
\multicolumn{7}{c}{2nd convolutional layer from the last} \\
\hline
CB & 0.9977 & 0.0011 & 0.7035 & 0.9998 & 0.0 & 0.6106 \\
SP & 0.9976 & 0.0011 & 0.5962 & 0.9948 & 0.0011 & 0.4923 \\
WB & 0.9976 & 0.0126 & 0.8544 & 0.9998 & 0.0267 & 0.8085 \\
\hline
\multicolumn{7}{c}{3rd convolutional layer from the last} \\
\hline
CB & 0.9888 & 0.0023 & 0.6117 & 0.9997 & 0.0 & 0.5336 \\
SP & 0.9941 & 0.0081 & 0.6266 & 0.9922 & 0.0046 & 0.3738 \\
WB & 0.9976 & 0.0207 & 0.7857 & 0.9997 & 0.0500 & 0.7049 \\
\hline
\multicolumn{7}{c}{4th convolutional layer from the last} \\
\hline
CB & 0.9888 & 0.0 & 0.5349 & 0.9997 & 0.0 & 0.4926 \\
SP & 0.9988 & 0.0302 & 0.6248 & 0.9976 & 0.0058 & 0.3705 \\
WB & 0.9952 & 0.0288 & 0.7581 & 0.9997 & 0.0848 & 0.6351 \\
\bottomrule
\end{tabular}
\caption{TPR, FPR and SIA for our defense against all the created attacks on different DNN internal layers.}
\label{tab:4}
\end{table}

\subsection{Experimental Results on Other Datasets} \label{sec:other_datasets}
We also evaluated our detector on datasets including PubFig, MNIST and F-MNIST. 
For PubFig, we randomly choose 20 classes, each with 80 training samples and 20 test samples. We arbitrarily select class 19 as the target class and embed a backdoor trigger in 2 training samples from each class except for the target class. The backdoor patterns considered for PubFig are trojan square (\textbf{SQ}) and trojan watermark (\textbf{WM})\cite{Trojan}, and the poisoned training set is then used for training a DNN with VGG-16\cite{VGG} architecture. We reserve 5 test images per class for detection. For each of MNIST and F-MNIST, we consider two attacks with backdoor patterns CB and WB, respectively, and the same attack settings for the multi-class attacks described in Sec.\ref{sec:exp_setup}. For each attack, a DNN with LeNet-5\cite{LeNet} architecture is trained on the poisoned training set. As shown in Table~\ref{tab:5}, our defender achieves similarly good performance on these datasets as for the CIFAR-10 dataset.
\begin{table}[t!] 
\centering
\small
\begin{tabular}{lllllll}
\toprule
\multirow{2}{*}{} & \multicolumn{2}{c}{PubFig} & \multicolumn{2}{c}{MNIST} & \multicolumn{2}{c}{F-MNIST} \\
\cline{2-7}
& SQ & WM & CB & WB & CB & WB \\
\hline
TPR & 0.9856 & 1.0    & 1.0    & 1.0    & 1.0    & 0.9951 \\
FPR & 0.1428 & 0.1428 & 0.0033 & 0.0022 & 0.0023 & 0.0127 \\ 
SIA & 0.7194 & 0.5507 & 0.9413 & 0.9764 & 0.6948 & 0.8039 \\ 
\bottomrule
\end{tabular}
\caption{TPR, FPR and SIA for our in-flight backdoor detector on datasets PubFig, MNIST and F-MNIST.}
\label{tab:5}
\end{table}

\section{Conclusion}
\label{sec:conclusion}

We proposed an in-flight defense that detects test images containing a backdoor trigger and infers the class of origin for each detected image. Our defense does not need access to the DNN's training set and is more efficient compared with existing in-flight defenses. We show the effectiveness of our detector experimentally for a wide variety of DNN architectures, datasets, and backdoor attack configurations.

\bibliographystyle{plain}

\end{document}